\documentstyle{article}
\begin{document}

\newcommand{\chand}{\sqcap} 

\newcommand{\pand}{\wedge} 

\newcommand{\sand}{\hspace{2pt}\mbox{\small \raisebox{0.0cm}{$\bigtriangleup$}}\hspace{2pt}} 

\newcommand{\chor}{\sqcup} 

\newcommand{\por}{\vee} 

\newcommand{\sor}{\hspace{2pt}\mbox{\small \raisebox{0.049cm}{$\bigtriangledown$}}\hspace{2pt}} 

\newcommand{\chall}{\hspace{1pt}\mbox{\Large $\sqcap$}} 

\newcommand{\chexists}{\hspace{1pt}\mbox{\Large $\sqcup$}} 

\newcommand{\pimplication}{\rightarrow} 

\newcommand{\precurrence}{\hspace{1pt}\mbox{\raisebox{-0.01cm}{\scriptsize $\wedge$}\hspace{-4pt}\raisebox{0.16cm}{\tiny $\mid$}}\hspace{2pt}} 

\newcommand{\srecurrence}{\mbox{\raisebox{-0.07cm}{\scriptsize $-$}\hspace{-5.9pt}\mbox{\raisebox{-0.01cm}{\scriptsize $\wedge$}\hspace{-4pt}\raisebox{0.16cm}{\tiny $\mid$}}}\hspace{2pt}} 

\newcommand{\coprecurrence}{\hspace{1pt}\mbox{\raisebox{0.12cm}{\scriptsize $\vee$}\hspace{-3.8pt}\raisebox{0.02cm}{\tiny $\mid$}}\hspace{2pt}} 

\newcommand{\cosrecurrence}{\mbox{\raisebox{0.20cm}{\scriptsize $-$}\hspace{-5.9pt}\mbox{\raisebox{0.12cm}{\scriptsize $\vee$}\hspace{-3.8pt}\raisebox{0.02cm}{\tiny $\mid$}}}\hspace{2pt}} 

\newcommand{\primplication}{\hspace{2pt}\mbox{\raisebox{0.033cm}{\tiny $>$}\hspace{-0.18cm} \raisebox{-0.043cm}{\large --}}\hspace{2pt}} 


\newcommand{\intimpl}{\Rightarrow}
\newcommand{\rneg}{\neg}               
\newcommand{\pneg}{\neg}               
\newcommand{\emptyrun}{\langle\rangle} 
\newcommand{\elzi}[1]{\mbox{\scriptsize $\parallel\hspace{-3pt} #1 \hspace{-3pt}\parallel$}}
\newcommand{\mleq}{\leftrightarrow}
\newcommand{\seqc}{\bigtriangleup}
\newcommand{\seqd}{\bigtriangledown}
\newcommand{\pst}{\Uparrow}
\newcommand{\pcost}{\Downarrow}
\newcommand{\seqst}{\uparrow}
\newcommand{\seqcost}{\downarrow}
\newcommand{\oo}{\bot}            
\newcommand{\pp}{\top}            
\newcommand{\xx}{\wp}               

\newcommand{\leg}[2]{\mbox{\bf Lm}^{#1}_{#2}} 
\newcommand{\win}[2]{\mbox{\bf Wn}^{#1}_{#2}} 
\newcommand{\legal}[2]{\mbox{\bf Lr}^{#1}_{#2}} 

\newcommand{\constants}{\mbox{\bf Constants}} 
\newcommand{\variables}{\mbox{\bf Variables}}    
\newcommand{\moves}{\mbox{\bf Moves}}        

\newcommand{\seq}[1]{\langle #1 \rangle}           
\newcommand{\tuple}[2]{\mbox{$ {#1}_{1},\ldots,{#1}_{#2}$}} 


\newcommand{\gneg}{\neg}                  
\newcommand{\mli}{\rightarrow}                     
\newcommand{\cla}{\mbox{\large $\forall$}}      
\newcommand{\cle}{\mbox{\large $\exists$}}        
\newcommand{\mld}{\vee}    
\newcommand{\mlc}{\wedge}  
\newcommand{\ade}{\mbox{\Large $\sqcup$}}      
\newcommand{\ada}{\mbox{\Large $\sqcap$}}      
\newcommand{\add}{\sqcup}                      
\newcommand{\adc}{\sqcap}                      
\newcommand{\adei}{\mbox{$\sqcup$}}      
\newcommand{\adai}{\mbox{$\sqcap$}}      
\newcommand{\clei}{\exists}
\newcommand{\tlg}{\bot}               
\newcommand{\twg}{\top}               

\newtheorem{theoremm}{Theorem}[section]
\newtheorem{definitionn}[theoremm]{Definition}
\newtheorem{problemm}[theoremm]{Problem}
\newtheorem{conjecturee}[theoremm]{Conjecture}
\newtheorem{claimm}[theoremm]{Claim}

\newtheorem{lemmaa}[theoremm]{Lemma}
\newtheorem{factt}[theoremm]{Fact}
\newtheorem{termnott}[theoremm]{Terminology and notation}
\newtheorem{termm}[theoremm]{Terminology}
\newtheorem{notationn}[theoremm]{Notation}
\newtheorem{corollaryy}[theoremm]{Corollary}
\newtheorem{conventionn}[theoremm]{Convention}
\newtheorem{example}[theoremm]{Example}
\newtheorem{remarkk}[theoremm]{Remark}
\newenvironment{definition}{\begin{definitionn} \em}{ \end{definitionn}}
\newenvironment{theorem}{\begin{theoremm} \em}{\end{theoremm}}
\newenvironment{lemma}{\begin{lemmaa} \em}{\end{lemmaa}}
\newenvironment{fact}{\begin{factt} \em}{\end{factt}}
\newenvironment{problem}{\begin{problemm} \em}{\end{problemm}}
\newenvironment{conjecture}{\begin{conjecturee} \em}{\end{conjecturee}}
\newenvironment{claim}{\begin{claimm} \em}{\end{claimm}}
\newenvironment{termnot}{\begin{termnott} \em}{\end{termnott}}
\newenvironment{term}{\begin{termm} \em}{\end{termm}}
\newenvironment{notation}{\begin{notationn} \em}{\end{notationn}}
\newenvironment{corollary}{\begin{corollaryy} \em}{\end{corollaryy}}
\newenvironment{convention}{\begin{conventionn} \em}{\end{conventionn}}
\newenvironment{remark}{\begin{remarkk} \em}{\end{remarkk}}
\newenvironment{proof}{{\bf PROOF}}{\ $\Box$ \ \\}

\title{Thoughts on sub-Turing interactive computability}
\author{Giorgi Japaridze
  \\  
 \\ Department of Computing Sciences, Villanova University, USA\\
 Email: giorgi.japaridze@villanova.edu\\
 URL: http://www.csc.villanova.edu/$^\sim$japaridz/
}
\date{}
\maketitle

\begin{abstract}  
The article contains an outline of a possible new direction for Computability Logic, focused on computability without infinite memory or other impossible-to-possess computational resources. The new approach would see such resources as external rather than internal to computing devices. They could or should be accounted for explicitly in the antecedents of logical formulas expressing computational problems.
\end{abstract}



\ 

\

This contribution is dedicated to my dear old friend and colleague, Vladimir Shalak, in celebration of his 70th birthday. His wisdom and optimism have always been an inspiration to all around him. I wish they continue to be so for many years to come. 

\section{Introduction}

The present article lies within the framework of {\em computability logic} (CoL) --- an ambitious long-term research project with a beginning but no end, initiated by the author in  \cite{Jap03} and actively pursued since then \cite{fund,bauerTOCL,kwon,Ver2010,Qu,xuIGPL}. Among many characterizations of CoL's formalism would be to say that it potentially provides a medium for communication between humans and computers, something inbetween programming languages on one hand, and 
the language used by humans  in their intellectual activities on the other hand. Just like the former, the language of CoL is formal and thus well understood by machines; and, just like the latter, it can be relatively easily   ``spoken'' by humans without any special expertise and training generally required for programmers.  

The question `what can be computed?' is fundamental to computer science. CoL is about answering this question in a systematic way using logical formalism, with formulas understood as computational problems and logical operators as operations on them. The first basic issue to be clarified here is 
what a {\em computational problem} means. With a few exceptions in the literature, starting from Turing \cite{Tur36}, this term usually refers to an entity 
that is modeled by a very simple interface between a computing agent and its environment, consisting in asking a question (input) and generating an answer (output). In other words, computational problems are understood as functions. This understanding, however, captures only a small part of our broader intuition and the reality of computational problems. Most  tasks that real computers perform are {\em interactive} and not reducible to simple pairs of 
input/output events. In such tasks, input/output events, also called {\em observable actions} \cite{goldin} by the computing agent 
and its environment, can be multiple and interspersed, perhaps taking place throughout the entire process of computation rather than just at the beginning (input) and the end (output) of it. Computability that CoL deals with is interactive computability,
and throughout this article by a ``(computational) problem" we always mean an interactive computational problem. This concept is formalized in Section 2 below. Section 3 provides a brief informal overview of operations generating complex computational problems from simpler ones, and Section 4 discusses the basic model of interactive computation used in CoL. These sections serve the purpose of establishing a background necessary for understanding the final Section 5, where the reader will find a discussion of a possible new direction into which CoL may branch: a direction that switches the attention of CoL from computability-in-principle (Turing computability) to sub-Turing comutability, where the latter  does not assume the presence of (in fact) supernatural resources such as the infinite-capacity tape memory. 

\section{Computational problems as games}
CoL starts by formalizing our broadest intuition of (interactive) computational problems. Technically 
this is done in terms of games. Namely, 
computational problems are understood as games between two agents: the {\em machine} and the {\em environment}, symbolically named as $\pp$ and $\oo$, respectively. Machine, as its name suggests, is a mechanical 
device with fully determined, algorithmic behavior, while the behavior of the environment, that represents a capricious 
user or the blind forces of nature, is allowed to be arbitrary. Observable actions by these two agents translate into 
game-theoretic terms as their {\em moves}, and interaction histories as {\em runs}.

Formally, a {\bf move} can be understood as any finite string over the standard keyboard alphabet. 
A move prefixed with $\pp$ or $\oo$ is called a {\bf labeled move} (with its label indicating who has made the move). A 
{\bf run} is a (finite or infinite) sequence of labeled moves, and a {\bf position} is a finite run. 

A  {\bf game} is defined as a pair $A=(\legal{A}{},\win{A}{})$, where:
\begin{itemize}
\item $\legal{A}{}$ is a set of runs such that $\Gamma\in\legal{A}{}$ iff all nonempty initial segments of $\Gamma$ 
are in $\legal{A}{}$; 
\item $\win{A}{}$ is a function of the type $\legal{A}{}\rightarrow\{\pp,\oo\}$.
\end{itemize}

The intuitive meaning of the $\legal{}{}$ component is that it tells us what the {\bf legal runs} of the game
are.
Extending the meaning of the term ``legal" to (labeled) moves, a move $\alpha$ is considered legal by a given
player $\xx$ in a given position iff adding $\xx\alpha$ to that position results in a legal position.    
The meaning of the $\win{}{}$ component is that it tells us who has won 
a given legal run. This function is usually extended  to all runs by
stipulating that an illegal run is always lost by the player who made the first illegal move.   

One of the main novel and distinguishing features of CoL's games among the other concepts of games studied in the logical literature  is the absence of  {\em procedural rules}, i.e., rules strictly regulating who and when can or should move, the most standard procedural rule being the one according to which the players should take turns in alternating order. Procedural rules usually yield the sort of games 
that CoL calls {\bf strict} --- games where in each position the player who has to move is strictly determined. CoL's 
games are not strict: it is possible that in a given position both players have legal moves; furthermore, having legal moves does not obligate the player to move, and in many cases it may have to decide not only {\em which} of the available moves to make, but also {\em whether} to make a move at all or rather wait to see how the adversary acts; watching the adversary's behavior while simultaneously thinking on its own decisions is also among the possible scenarios. How, exactly, runs are generated will be seen in Section 4. For now, we can adopt an informal yet perfectly accurate explanation, according to which either player is free to make any move at any time. Based on this, our games 
can be called {\bf free}. Strict games are not powerful enough to adequately capture all reasonable interactive tasks. 
Most of such tasks that we perform in everyday life are free rather than strict, and so are the majority of computer communication/interaction protocols. CoL's free games are most general 
of all (two-player, two-outcome) games, which which makes them most natural, as well as powerful, modeling tools for interactive tasks. Strict games can be thought of as a special case of CoL's free games where the $\legal{}{}$ component is such that, in every position, at most one of the two players has legal moves.

The class of  free games in the above-defined sense is, however, too general. There are games where the chances of the machine to succeed may essentially depend on the relative speed at which its adversary responds, 
and we do not want to consider that sort of games as meaningful computational problems. A simple example would be the game where all moves are legal and that is won by the player who moves first. This is merely a contest of speed. Furthermore, if the interaction happens by exchanging
message through a (typically asynchronous) network, it can also be a contest of luck, where success may as well
depend 
on whose message was delivered to the arbiter sooner. Below we define a subclass of games called {\em static games}.
Intuitively, static games are games where speed is irrelevant: in order to succeed, it only matters {\em what} to do 
(strategy) rather than {\em how fast} to do (speed). One of the main theses\label{thes} on which the philosophy of CoL relies is that the concept of static games is an adequate 
formal counterpart of our intuitive notion of ``pure", speed-independent computational problems. \cite{Jap03} contains
a detailed discussion in support of this thesis. Here we only give a formal definition.

For $\xx\in\{\pp,\oo\}$, we say that a run $\Omega$ is a {\bf $\xx$-delay} of a run $\Gamma$ iff the following two conditions are satisfied:
\begin{itemize}
\item For each player $\xx'$, the subsequence of $\xx'$-labeled moves of $\Omega$ is the same as that of $\Gamma$;
\item For any $n,k$, if the $n$th $\xx$-labeled move is made later than (is to the right of) the $k$th non-$\xx$-labeled move in $\Gamma$, then so is it in $\Omega$.
\end{itemize}
Intuitively this means that in  $\Omega$  each player has made the same sequence of moves as in $\Gamma$, only, in $\Omega$, $\xx$ might have been acting with some delay. 

Now,  we say that 
a game  $A$ is {\bf static} iff, whenever $\win{A}{}(\Gamma)=\xx$ and $\Omega$ is a $\xx$-delay of 
$\Gamma$, we have $\win{A}{}(\Omega)=\xx$.

At this point we are ready to formally answer the question regarding what CoL exactly means by computational problems:  {\em an  
{\bf interactive computational problem (ICP)} is a static game}.   

\section{Game operations}

The logical connectives and quantifiers of CoL represent operations on ICPs, some of which take inspiration in game operations studied by Lorenzen \cite{Lor59,Fel85}, Hintikka \cite{Hin73} and 
(especially) Blass \cite{Bla72,Bla92}, and resembling certain operators of linear logic \cite{Gir87}.  Over three dozen ICP operations have been officially defined and studied so far, and more are probably yet to come due to the open-ended character of the formalism of CoL. The operations  include several natural sorts of negation, disjunction/conjunction, implication, quantifiers. There are also several sorts of unary operations, called {\em recurrences}, with no analogs in classical logic. Below we briefly and informally survey only a few inhabitants of the operation zoo of CoL to get some intuitive feel for the approach. See \cite{fund} for a comprehensive account and formal definitions.  

The (basic) {\bf negation} $\gneg A$ of a game $A$ is defined as $A$ with the roles of the two players switched: $\pp$'s legal moves and wins become $\oo$'s legal moves and wins, and vice versa. E.g., if {\em Chess} is the game of chess from the point of view of the white player, then $\gneg${\em Chess} will be the same game from the point of view of the black player.

The  operations $\adc$ ({\bf chand}, abbreviting ``choice AND''), $\add$ ({\bf chor}), $\ada$ ({\bf chall}) and $\ade$ ({\bf chexists}) can be seen as ``constructive''  
 versions of conjunction, disjunction, universal quantifier and existential quantifier, respectively. 
$\ada xA(x)$ is the game where, in the initial position, only $\oo$ has legal moves. Such a move consists in 
a choice of one of the elements $c$ of the universe of discourse. After $\oo$ makes such a move $c$, 
the game continues (and the winner is determined) according to the rules of $A(c)$. If there was no initial move, $\pp$ is considered the 
winner as there was no particular instance of the problem  specified by $\oo$ that $\pp$ failed to solve.
$A\adc B$ is similar, 
only here the choice is just made between ``0" ({\em left}) and ``1" ({\em right}). $\ade$ and $\add$ are 
symmetric to $\ada$ and $\adc$, with
the only difference that now it is $\pp$ rather than $\oo$ who makes the initial move/choice. An example would help. The problem of computing function $g$ can be specified as 
\(\ada x\ade y\bigl(g(x)=y\bigr).\)  This is a two-move-deep game where the first legal move --- selecting a particular 
value $k$ for $x$ --- is by $\oo$, which brings the game down to $\ade y(g(k)=y)$.  The second move --- selecting a value $m$ for $y$ --- is by $\pp$, after which the game continues (or rather ends) as $g(k)=m$. 
Such a run $\seq{\oo k,\pp m}$ is then considered won by $\pp$ iff
$m$ really equals $g(k)$. Obviously $g$ is computable in the standard sense iff there is a machine that wins
$\ada x\ade y\bigl( g(x)=y\bigr)$ against any possible (behavior of the) environment. 

The operations $\mlc$ ({\bf pand}, abbreviating ``parallel and'') and $\mld$ ({\bf por})  generate parallel combinations of games. Playing $A_0\mlc A_1$ or $A_0\mld A_1$ means playing the two games $A_0$ and $A_1$ concurrently.  
Every legal run of such a game will thus consist of two --- arbitrarily interspersed --- subruns, with one subrun being a legal run of $A_0$,
 and the other subrun a legal run of $A_1$. 
Both $A_0\mlc A_1$ and $A_0\mld A_1$ thus have exactly the same $\legal{}{}$ component. Only their $\win{}{}$ components are different: in order for $\top$ to win, in $A_0\mlc A_1$ it needs to win in both components, while in the case of $A_0\mld A_1$ 
winning in one of the components is  sufficient. To appreciate the difference between the choice and the parallel sorts of operations, compare the games {\em Chess}\hspace{1pt}$\mld\gneg${\em Chess} and {\em Chess}\hspace{1pt}$\add\gneg${\em Chess}. The former is 
a play on two boards, which is very easy for an agent to win even against Kasparov himself: all that would suffice to do is
to copy in one conjunct the moves made by the adversary in the other conjunct, and vice versa. On the other hand, winning {\em Chess}$\hspace{1pt}\add\gneg${\em Chess} against Kasparov, where from the very beginning the agent has to select one of the disjuncts and then successfully play the chosen one-board game, is not easy at all.

Among the most important operations is {\bf pimplication} (parallel implication) $\mli$. $A\mli B$ is formally defined as $\gneg A\mld B$. Intuitively, this is the problem of {\bf reducing} $B$ to $A$: solving $A\mli B$ means solving $B$ having $A$ as 
an (external) {\em computational resource}. Generally, computational resources are symmetric to computational tasks/problems: what is a problem for one player to solve, is a resource for the other player to use, and vice versa. 
Since in the antecedent of $A\mli B$ the roles are switched, $A$ becomes a problem for $\oo$ 
to solve, and hence a resource that $\pp$ can use. Thus, CoL's semantics of computational problems is, at the same time, a semantics of computational resources, which offers a materialization of the (so far rather abstract) resource philosophy associated with linear logic. 
To get a feel of $\mli$ as a problem reduction operation, let us consider  reduction of the acceptance problem to the halting problem. The halting problem   can be expressed by
$\ada x\ada y \bigl(\mbox{\em Halts}(x,y) \add \gneg \mbox{\em Halts}(x,y)\bigr)$,
where $\mbox{\em Halts}(x,y)$ is the predicate 
``Turing machine  $x$ halts on input $y$". Similarly, the acceptance problem  (in the sense of \cite{sipser}) can be expressed by
$\ada x\ada y \bigl(\mbox{\em Accepts}(x,y) \add \gneg \mbox{\em Accepts}(x,y)\bigr)$. 
 While the acceptance problem is not decidable, it is algorithmically reducible to the halting problem. In our terms this means nothing but that there is a machine that always wins 
 \[\ada x\ada y \bigl(\mbox{\em Halts}(x,y) \add \gneg \mbox{\em Halts}(x,y)\bigr)\mli \ada x\ada y \bigl(\mbox{\em Accepts}(x,y) \add \gneg \mbox{\em Accepts}(x,y)\bigr).\]
A strategy for solving this problem is to wait till the environment specifies values $k$ and $n$ for $x$ and $y$ in the consequent, then select the same values $k$ and $n$ for  $x$ and $y$ in
the antecedent (where the roles of the machine and the environment are switched), and see whether $\oo$ responds by {\em left} or {\em right} there. If the response is 
{\em left}, simulate machine $k$ on input $n$ until it halts, and select, in the consequent, {\em left} or {\em right} depending on whether the simulation accepted or rejected. And if $\oo$'s response in the antecedent is {\em right}, then select {\em right} in the consequent. 
We can see that what the machine did in the above strategy indeed was reduction: it used an (external) solution 
to the halting problem to solve the acceptance problem. 

The next operation on our list is {\bf precurrence} (parallel recurrence) $\precurrence$. Intuitively, $\precurrence A$ is nothing but the infinite $\mlc$-conjunction $A\mlc A
\mlc A\mlc A\mlc\ldots$. That is, as a resource, $\precurrence A$ is $A$ that can be used and reused arbitrarily many times.  In terms of $\precurrence$ we can define the {\bf primplication} ({\bf p}arrallel-{\bf r}ecurrence-based {\bf implication})   $\primplication$ by $A\primplication B\ =\ \precurrence A\mli B$. 
The difference between $A\mli B$ and $A\primplication B$ is that while in the former every act of resource ($A$) consumption
is strictly accounted for, $A\primplication B$ allows unlimited usage of resources. This makes computability of $A\primplication B$ a generalization of Turing reducibility from simple, two-step problems to all ICPs. To get a feel of $\primplication$, remember the Kolmogorov complexity problem. It can be expressed by  
\(\ada t\ade z\hspace{2pt}\mbox{\em K}\hspace{2pt}(z,t),\)
where  $\mbox{\em K}\hspace{2pt}(z,t)$ is the predicate ``$z$ is the smallest (code of a) Turing machine that returns $t$ on input $0$".
Having no algorithmic solution, the Kolmogorov complexity problem, however, is Turing-reducible to the halting problem. In our terms, this means nothing but that there is a machine that always wins the game
\[\ada x\ada y \bigl(\mbox{\em Halts}(x,y) \add \gneg {\em Halts}(x,y)\bigr)\ \primplication\ \ada t\ade z\hspace{2pt} \mbox{\em K}\hspace{2pt}(z,t).\]
Here is a strategy for such a machine: Wait till the environment selects a particular value $m$ for $t$  in the consequent. 
Then, starting from $i=0$, do the following. Make a move by specifying $x$ and $y$ as $i$ and $0$, respectively, in the $i$th copy ($\mlc$-conjunct) of the  antecedent. If the environment responds by {\em right} there,
increment $i$ by one and repeat the step; if the environment responds by {\em left}, simulate machine $i$ on input $0$ until it halts; if you see that  machine $i$ returned $m$, make a move in the consequent by specifying $z$ as $i$; otherwise, increment $i$ by one and repeat the step.
We have just seen an example of a weak ($\primplication$) reduction to the halting problem, which cannot be replaced with the strong ($\mli$) reduction.  

Here comes the third, sequential, sort of conjunction $\sand$ ({\bf sand}) and disjunction $\sor$ ({\bf sor}). The latter can be defined in terms of the former by $A\sor B=\neg(\neg A\sand \neg B)$.   The meaning of the game $A\sand B$ is similar to that of $A\adc B$, with the difference that while in the latter 
$\oo$ has to choose one of the two conjuncts at the very beginning of the play, in the former this choice can be made 
at any later time: a legal run of $A\seqc B$ starts and proceeds in its default $A$ component. If at some point $\oo$ decides to switch components, $A$ will be abandoned without the possibility to come back to it later, and the (rest of the) play 
will restart and continue as a play of $B$. Let us take a look at an example. 
Obviously   
decidability of a predicate $A(x)$ means nothing but computability of $\ada x(\gneg A(x)\add A(x))$. Replacing $\add$ by $\sor$ in this formula makes it express recursive enumerability of $A(x)$. Indeed, when $A(x)$ is r.e., the following strategy is obviously successful for $\ada x(\gneg A(x)\sor A(x))$: after $\oo$ specifies a value $m$ for $x$, start looking for a witness for $A(m)$; make a component-switch move 
if and when such a witness is found. On the other hand, with some additional thought, it would not be hard to see that no machine can win $\ada x(\gneg A(x)\sor A(x))$ when $A(x)$ is not r.e..

Finally, the sequential recurrence ({\bf srecurrence}) $\srecurrence A$  of $A$ is nothing but the infinite conjunction  $A\seqc A\seqc A\seqc\ldots$ (one technical detail that needs to be clarified here: if a switching move is made infinitely many times in a legal run of $\srecurrence  A$, $\pp$ is considered the winner).  
\section{Interactive machines}
Now that we know what interactive computational problems are, it is time to explain what their computability means.
As we remember, the central point of CoL's philosophy is to require that agent $\pp$ be implementable as a computer program, with effective and fully determined behavior. On the other hand, the behavior (including speed) of agent $\oo$ can be arbitrary. This intuition is captured by the model of interactive computation where $\pp$ is formalized as what CoL, for historical reasons,   calls {\bf HPM} (Hard-Play Machine).  
Below is only an informal description of this model. The omitted and not very interesting technical details can be 
very easily restored by anyone familiar with Turing machines. And, just like Turing machines, our play machines are 
highly robust with respect of all sorts of reasonable variations of those details. 

An HPM a Turing machine (TM) $\cal M$ with the capability of making moves. Along with the ordinary read/write {\em work tape}, such a machine has an additional read-only tape called the  
{\em run tape}. The latter serves as a dynamic input, spelling, at any time, the current position of the game.
Every time one of the players makes a move, that move (with the corresponding label) is automatically appended to the contents of the run tape. There are no limitations to the frequency at which the environment can make moves, so on every transition from one computation step to another, any finite number of new $\oo$-labeled moves  may appear on the run tape. The technical details about how exactly $\cal M$ makes moves are not interesting and 
can be specified in various equivalent ways. For clarity let us say that this happens by writing the move in a certain special segment of the work tape and then entering 
one of the specially designated states called {\em move states}.
  
The sequence of computation steps (configurations) obtained this way forms what we call a {\em computation branch} of $\cal M$. The latter incrementally spells a run on the run tape in the obvious sense.  We call such a  run a {\em run generated by $\cal M$}. The set of all such runs corresponds to the set of all possible behaviors by the environment. Then,
for an ICP $A$, we say that ${\cal M}$ {\bf computes (solves, wins)} $A$ iff, for every run $\Gamma$ generated by $\cal M$, 
$\win{A}{}(\Gamma)=\pp$.  Finally, we say that $A$ is 
{\bf computable (solvable, winnable)} iff there is an HPM that computes $A$. 

{\bf Validity} in CoL can now be defined  by saying that a formula $F$ is valid iff there is an HPM $\cal M$ such that $\cal M$ wins $F$ no matter what particular ICPs its atoms stand for. $P\mld\neg P$ is an example of a valid formula, and $P\add\neg P$ an example of an invalid formula.

Philosophically, CoL is based on the following two major theses:

{\bf Thesis 1.} Our intuitive concept of ``pure", speed-independent interactive computational problems is adequately captured by the formal concept of static games (see page \pageref{thes}). 

{\bf Thesis 2.} Our intuitive concept of computability of such problems is adequately captured by the above formal concept of computability (by HPMs). 

Thesis 2 is thus a generalization of the Church-Turing thesis \cite{church} to interactive 
computability. Anyone critical or skeptical about CoL's approach, would have to either argue that interactive computability is not worth
studying, or argue that the approach is not really about interactive computability. In the latter case, as a minimum, the skeptic would be expected to make a good case against Theses 1 and 2.

\section{From Turing to sub-Turing computability}  
According to the Church-Turing thesis 
and the above more general Thesis 2, TMs and HPMs present adequate universal models of mechanical computing. Both of these 
models, however, assume the presence of an infinite work tape --- an abstraction whose legitimacy might be philosophically 
questioned. Neither TMs nor HPMs can physically exist for the simple reason that no real mechanical device will ever have an infinite (even if only potentially so) internal memory. The reason why such an abstraction still does not cause 
frustration  
is that the tape can be easily thought of as an {\em external resource}, and TMs or HPMs identified only with their finite control part; then and only then, they indeed become implementable devices. Yet, the formal treatment of TMs or 
HPMs does not account for this implicit intuition, and views the infinite work tape as a part of the machine --- after all, having such a tape is basically what distinguishes TMs from finite automata (DFAs) only capable of handling regular languages/problems.  

The idea that one can pursue is: {\em Deprive HPMs of the (infinite) internal work tape and make them truly finite devices.} Let us call the resulting machine model {\bf HPM}$^*$.  Replacing HPM by HPM$^*$ yields  new versions of some of CoL's old basic concepts; we will be using the same superscript $^*$ --- as in ``CoL$^*$", ``computability$^*$", ``reducibility$^*$ etc. --- to differentiate those new concepts from their old counterparts. The above idea would not be very appealing within the framework of traditional approaches, for we would end up with the class of computable$^*$ problems   
narrowed down to regular or, at best (with multiple, bidirectional input read heads), logarithmic-space computable ones. 
In our case, however, such a step is perfectly meaningful and even imperative. The unlimited computational power/resource 
that TMs or HPMs posses --- which lies in their infinite memory --- can be formalized as a certain game $T$ such that computability of a problem $A$ in the old sense now will mean nothing but its reducibility$^*$ to $T$, i.e. computability$^*$  of the problem $T\mli A$. Thus, our step from CoL to CoL$^*$ 
signifies no longer taking for granted the unlimited power of TMs, and explicitly turning   
that power (if and when its presence needs to be assumed) into an external computational resource, moving it from the work tape to the run (input) tape. 

Imagining how to formalize $T$ as an ICP would not be hard. One of the ways is to express it in terms of the simpler game 
$BIT$ that represents a one-bit, write-once storage device. $BIT$, which can be treated as a logical constant in the language of CoL$^*$, is the finite game where the first legal move --- 0 or 1 ---  is by $\oo$. The meaning of such a move is writing (to be remembered) the corresponding bit. The second  --- empty --- move  is also by $\oo$, and it means requesting a read. 
The final move is by $\pp$, who should 
return the very bit that was written.  
To turn {\em BIT} into 
rewritable one-bit memory, we prefix it with the operator $\srecurrence$. Finally, to obtain infinite (in particular, random-access) memory, we further prefix all that by $\precurrence$. Thus, $T$ can be defined as $\precurrence\srecurrence${\em BIT}. Considering this resource in the context 
of CoL's old approach would not make any sense because obviously $T$, just like all computable problems, is equivalent to $\pp$.
In the new context of CoL$^*$ this equivalence, however, obviously ends: $T$ is computable but (unlike $\twg$) not computable$^*$. 

Computing$^*$ $A$ 
and computing$^*$ $T\mli A$ correspond to the intuitions of computing $A$ in the strongest and the weakest sense, respectively, based on which we get interactive 
generalizations of the concepts of regular and decidable languages. Between these two extremes is a whole spectrum of intermediate versions of limited-resource ({\em sub-Turing}) computabilities, that can be expressed by replacing $T$ with weaker$^*$ ICPs. 
For instance, $STACK(BIT)\mli A$ expresses computability of $A$ by interactive pushdown automata, where $STACK$ is a (yet another new)
recurrence operator that allows stack-style activation/usage of (multiple copies of) its argument. In order to capture 
space complexity and similar concepts, we would need to introduce {\em bounded versions} of our recurrence operations  --- versions that take a variable as an additional argument and thus are, in a sense, quantifiers. E.g., $\precurrence x A$ would be the version of $\precurrence A$ where 
only $x$ parallel copies of $A$ can be used. Then $\precurrence x\srecurrence BIT\mli A$ expresses a natural interactive generalization of the concept of computing $A$ in space $x$. One of the ways to capture time complexity is
through the new sort of quantifier-style operation $\tau$. Roughly, $\tau xA$ is the version of $A$ where at most 
$x$ moves are allowed to be made. One could make a case that then $\tau x T\mli A$ expresses what is an adequate interactive counterpart of the concept of computing $A$ in time $x$.      

The above discussion is just an abstract outline of the idea of CoL$^*$, serving the purpose of 
giving the reader some feel of its potential. Materializing this  half-baked idea requires  
working out many technical details (e.g., what an HPM$^*$ exactly is and how it accesses its run tape or an equivalent 
dynamic-input device --- details that 
were irrelevant in the context of CoL but begin to matter in CoL$^*$). This will be nontrivial yet apparently doable work. The interest toward this approach, which gives computability logic new dimensions and takes it to a new level of intellectual interest, is in its potential to mature into a comprehensive logical theory of computational resources. The old CoL already {\em is} such a theory, but it uses a very high level of abstraction in understanding computability and hence refuses to distinguish between different \mbox{sub-Turing} computable resources. 
CoL$^*$, on the other hand, should provide systematic means for studying computational resources/problems  at the finest level of abstraction, based on just one single model HPM$^*$ of computation as opposed to the confusing variety of models that the traditional automata theory and complexity theory have to deal with.

\end{document}